\documentclass[preprintnumbers,amsmath,amssymb,twocolumn]{revtex4}
\usepackage[colorlinks=true,citecolor=red,filecolor=green,linkcolor=blue,pdfnewwindow=true]{hyperref}
\usepackage{graphicx,ulem}
\usepackage{makeidx}
\usepackage[caption=false]{subfig}
\usepackage{bm}
\usepackage[title]{appendix}

\def\prl#1#2#3{{ Phys. Rev. Lett.} {\bf #1}, #2 (#3)}

\def\pre#1#2#3{Phys. Rev. E {\bf #1}, #2 (#3)}

\def\chaos#1#2#3{Chaos {\bf #1}, #2 (#3)}

\def\ep{\varepsilon}

\def\ie{i.e. }

\def\beqr{\begin{eqnarray}}
\def\eqnr{\end{eqnarray}}
\def\beq{\begin{equation}}
\def\bc{\begin{center}}
\def\ec{\end{center}}
\def\eqn{\end{equation}\noindent}
\begin{document}
\title{Occurrence and stability of chimera states in multivariable coupled flows}
\author{Anjuman Ara Khatun$^{1}$, Haider Hasan Jafri$^{1}$}
\affiliation{$^{1}$Department of Physics, Aligarh Muslim University, Aligarh 202 002, India}
\begin{abstract}
Study of collective phenomenon in populations of coupled oscillators are a subject of intense exploration in physical, biological, neuronal and social systems. Here we propose a  scheme for the creation of chimera states, namely the coexistence of distinct dynamical behaviors in an ensemble of multivariable coupled oscillatory systems and a novel scheme to study their stability. We target bifurcation parameters that can be tuned such that out of the two coupling parameters one pushes the system to synchrony while the other one takes it to the desynchronized state. The competition between these states result in a situation that a certain fraction of oscillators are synchronized while others are desynchronized thereby producing a mixed chimera state. Further changes in couplings can result in the desired form of the state which could be completely synchronized or desynchronized. Using the model example of coupled R\"ossler systems we show that their basin of attraction are either riddled or intertwined. We use Strength of incoherence and Master Stability function (MSF) as the order parameters to verify the stability of chimera states. MSF for different attractors are found to possess both negative and positive  values indicating the coexistence of stable synchronized dynamics with desynchronized state.    \end{abstract}
\maketitle

{\bf Dynamics of coupled oscillators show an intriguingly complex behavior. In classical setting, it was shown that a set of nonlocally coupled phase oscillators can coexist as synchronized and desynchronized group namely the chimera states \cite{strogatz}. This state was subsequently investigated for a variety of settings including topologies, couplings and oscillator types \cite{panaggio,tinsley,abrams1}.  Earlier studies have shown that coupled oscillators should have weak and nonlocal coupling to show chimera states. At other instances it was argued that for globally coupled oscillators chimera states can occur as a result of multistability \cite{chandrasekar,sangeeta1}. On changing the coupling parameter the transition to various collective states can be achieved depending upon initial conditions. In the present study, we couple the system in two variables to explore these states. Though chimera states can be achieved through coupling in single variable only, the coupling in second variable allows us to restrict the regions where chimera states can be avoided.  We also introduce the idea of Master Stability Function (MSF) \cite{pecora-msf} to characterize the chimera states by exploiting the fact that negativity of MSF implies that the synchronized states are stable. Since for multistable systems, the MSF is calculated for all the coexisting attractors \cite{bocaletti}, we show that for chimera states, MSF is positive for one set of attractors that are desynchronized while it is negative for the synchronized set.}

\begin{section}{Introduction}
Phenomenon of synchronization is of great  importance in various fields namely physics, chemistry, biology, and medicine. Numerous studies have been devoted to the transition from desynchronized to synchronized regimes \cite{pik,pecora}. Over the last decade there has been a resurgence of interest where ensembles of nonlocally coupled oscillatory units can show coexisting coherent and incoherent dynamics namely the {\it chimera states}. This has been first reported in an ensemble of coupled phase oscillators with nonlocal couplings \cite{kuramoto,strogatz}.  It was shown that an array of identical oscillators split into coherent and incoherent domains.  Chimera states have also been reported at other instances, namely, in chemical oscillators \cite{tinsley}, system with time delay \cite{sethia,oleh, sheeba1,sheeba2}. It was also demonstrated that the spiral wave chimeras can exist in dynamical systems under the influence of nonlocal coupling \cite{chad}.
A scheme for the mechanism of the coherence-incoherence transition in networks with nonlocal coupling of variable range was described in \cite{hagerstrom,scholl-prl,scholl-pre}. These states were also studied for a globally coupled network of semiconductor lasers with delayed optical feedback \cite{bohm}. Recently, it has been shown that the chimera states may emerge as a result of induced multistability due to the couplings that bring the system parameters in a multistable regime \cite{sangeeta1}. The coupling effectively changes the parameter of the system leading to multistable dynamics and finally the creation of chimera states.

 In this work we describe how chimera states emerge by tuning the coupling parameters in globally coupled oscillators. These oscillators are coupled in more than one variable. It has also been reported that in multivariable coupled oscillators there is an enhancement in the stability of complete synchronization of the oscillators. Consequently, the MSF was found to have lower values on tuning the coupling parameters \cite{escoboza}.  The coupling parameter corresponding to $y-$ variable is $\ep_1$ whereas $\ep_2$ denotes the coupling strength in $z-$ variable.  Keeping $\ep_2$ fixed and changing $\ep_1$ in a particular direction, one can create chimeras from an ensemble of desynchronized oscillators. The newly created chimera state can be destroyed by further changing $\ep_1$ results in a state where all the oscillators are synchronized. Thus one can create chimera states for a particular range of $\ep_1$, say $\Delta \ep_1$ which can be varied or controlled by tuning  $\ep_2$. Though this phenomenon can be observed by introducing the coupling in one variable only, the coupling in second variable is used such that the system moves to a point where one cannot observe mixed state for any value of the first coupling parameter. We see that with the introduction of coupling in the second variable $\ep_2$, one can essentially select the range of first parameter $\Delta \ep_1$ for which mixed state or chimeras may be observed.
 Thus, we are able to target regions in the parameter state such that a desired state is obtained or in other words we are able to force a given system such that it shows robustly a behavior that is apriori chosen. This techinque is fairly general in the sense that emergence of chimeric state is not a consequence of  time-delay \cite{sethia} or inherent multistability in the system \cite{sangeeta1}.

 This paper is organized as follows. In the following section, we describe the typical features of the system and couplings that are germane to the present study. We also validate them using quantitative measures. In Sec. \ref{basin1}, we explore the basin of attraction and Lyapunov exponents. In Sec. \ref{msf}, we study the Master Stability function to check the stability of the synchronized state. Finally, in Sec.\ref{summary} we summarise our findings accompanied by some discussion.

\end{section}

\section{The Parameter modulated R\"ossler oscillator }
\label{ros}

\begin{figure}
\centering
\scalebox{0.5}{\includegraphics{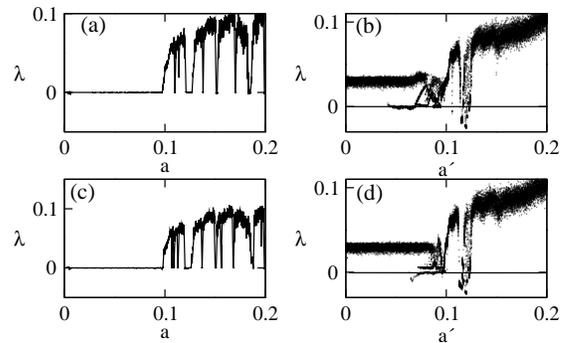}}
\caption{Variation of the Largest Lyapunov exponent for system given by Eq.~\ref{eq1} for the parameter values $b=0.1$ and $c=9$ when the oscillators are (a) uncoupled \ie at $\ep_1=\ep_2=0$ (b) coupled at $a=0.1$ and $\ep_2=0$ for $N=2$,  (c) uncoupled  for $c=9.1$ and $\ep_1=\ep_2=0$ and  (d) coupled for $N=2$ at $a=0.1$, $c=9.0$ and $\ep_2=0.1$. These plots have been generated over 30 different initial conditions.}
\label{fig1a}
\end{figure}

Consider an ensemble of $N$ globally coupled R\"ossler oscillator with linear diffusive coupling. The dynamical equations are given by
\beqr
\label{eq1}
\dot{x}_i&=& -y_i -z_i \nonumber \\
\dot{y}_i&=& x_i+ay_i+\ep_1\left(\frac{1}{N-1}\sum_{j\ne i}y_j-y_i\right) \nonumber \\
\dot{z}_i&=&b+z_i(x_i-c) +\ep_2\left(\frac{1}{N-1}\sum_{j\ne i}z_j-z_i\right)
\eqnr

where, we chose the system parameters to be $a=0.1$, $b=0.1$ and $c=9$ where the dynamics of uncoupled system is chaotic as shown in Fig.~\ref{fig1a} (a) \ie Lyapunov Exponent (LE) is positive. At this set of parameter values, there is no multistability in the uncoupled system. However, earlier studies have shown that two mutually coupled systems, which individually demonstrate Feigenbaum route to chaos, show multistability \cite{shabunin,vadivasova}.

These parameter values are chosen at a point that is very close to the bifurcation point that exists in the autonomous system. We use the coupling strengths $\ep_1$ and $\ep_2$ as control parameters for the creation and annihilation of the chimera states. The individual oscillator can be considered under the influence of the mean field given by $\bar{f_y}(t)=\frac{1}{N-1}\sum_{j\ne i}y_j$ and $\bar{f_z}(t)=\frac{1}{N-1}\sum_{j\ne i}z_j$ and one can consider the effectively driven system to be (c.f. Eq.~(\ref{eq1}))
\beqr
\dot{y}_i&=&x_i + a^\prime y_i + \ep_1\bar{f}_y(t) \nonumber \\
\dot{z}_i&=&b + z_i(x_i - c^\prime) + \ep_2\bar{f}_z(t) 
\eqnr

where, $a^\prime=a-\ep_1$ and $c^\prime=c+\ep_2$.

The reason that the ensemble of globally coupled oscillators go to chimera states can be understood by comparing the dynamics of two coupled oscillators under the influence of couplings of variable magnitude with that of the uncoupled oscillator. The variation of the Largest Lyapunov exponent for the autonomous system, w.r.t parameter $a$, is shown in Fig.~\ref{fig1a}(a). As discussed earlier, under the influence of coupling parameter $\ep_1$ at $\ep_2=0$, the parameter $a$ varies as $a^\prime=a-\ep_1$. As $\ep_1$ is reduced below zero, effective value of $a^\prime$ increases where LE is positive leading to desynchronized state and the dynamics corresponding to the same value of $a$ in the autonomous system is chaotic.  However, at the intermediate values of $\ep_1$ above zero, one can expect that the system will enter the mixed regime. If one observes the LE of the coupled system Fig.~\ref{fig1a}(b), we see that there is an induced multistability in the system just below $a^\prime=0.1$ \cite{shabunin,vadivasova}. This is the region where one can observe chimera states. Any increase in $\ep_2$ will result in an increase in the effective value of parameter $c$ ($c\rightarrow c^\prime =c+\ep_2$) where the dynamics stays chaotic and hence we expect desynchrony in the population. Putting, the two mechanisms together: decrease in $\ep_1$ causes population to synchronize whereas an increase in $\ep_2$ results in their desychronization.  

If one tunes the second coupling parameter $\ep_2$ (say $\ep_2=0.1$), one can observe that the multistable region has shifted.  The comparison of the dynamics of coupled system with that of the autonomous system at $c=9.1$ is shown in Figs.~\ref{fig1a}(c) and (d). Thus one can infer that one coupling parameter is pulling the system towards a region of synchrony and the other one towards desynchrony. This competition results in induced multistability in coupled system and hence gives rise to the chimera states. If the parameter that drags the system towards synchrony is increased then it results in a state where all the oscillators are synchronised which may be destroyed by changing the other coupling parameter.

 \begin{figure}
\centering
\scalebox{0.8}{\includegraphics[width = 1\columnwidth,angle=270]{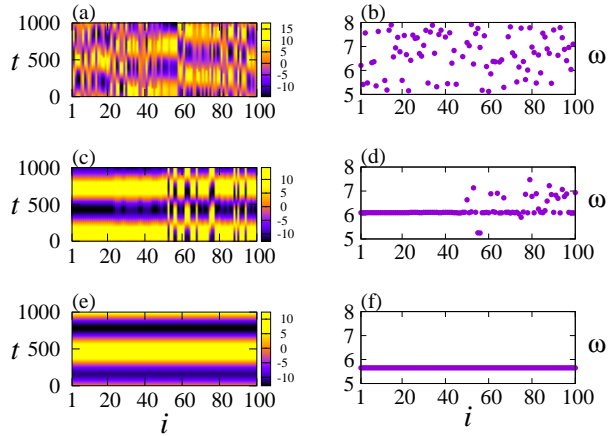}}
\caption{Dynamics for $N=100$  globally coupled oscillators (Eq.~\ref{eq1}) for the parameter values $a=b=0.1$, $c=9$ and $\ep_2=0.1$ where the left panel describes space-time plots and right panel their corresponding frequencies. (a) space-time plot and (b) frequencies for  $\ep_1=-0.08$ where all the oscillators are desynchronized. For chimera states (c) space-time and (d) frequencies are plotted for  $\ep_1=0.023$. At $\ep_1=0.07$ (e) space-time plot and (f) frequencies are shown where all the oscillators are synchronized. }
\label{fig2-new}
\end{figure}

 In order to realise this transition we start by fixing the coupling in $z$ variable \ie $\ep_2=0.1$ and changing the other coupling constant $\ep_1$ along a particular direction. It is observed that at $\ep_1=-0.08$  all the oscillators of the ensemble Eq.~(\ref{eq1}) are completely desynchronized as shown in Fig.~\ref{fig2-new}(a) that represents the time evolution of the $x$ variable.  This can be confirmed by looking at the frequencies plotted in Fig.~\ref{fig2-new}(b). At this coupling value the effective system parameter becomes $a^\prime=0.18$, where the Lyapunov exponent of the uncoupled system is positive. As we increase the coupling, \ie $\ep_1=0.023$ we observe that some oscillators go to synchronized state while others are desynchronized  as described by the time evolution of the x-variable described by  Fig.~\ref{fig2-new}(c) and frequency of the individual oscillator shown in Fig.~\ref{fig2-new}(d). This is the value for which the coupled oscillators exhibit multistabilty for $a^\prime=0.077$ (Fig.~\ref{fig1a} (b) and (d)) \cite{shabunin,vadivasova}.  Here, we see that there is a range of $\ep_1$ values, say $\Delta \ep_1$ for which the system exhibits chimera states for a given value of $\ep_2$. $\Delta \ep_1$ can be varied or controlled by changing $\ep_2$, \ie effectively shifting the system to a new value of parameter $c$ in the parameter space. Again, on increasing the coupling further, one can see that the coexisting coherent and incoherent states are destroyed and the system goes to a state where all the oscillators are synchronized  at $\ep_1=0.07$ as shown in Fig.~\ref{fig2-new}(e) and (f). Here $a^\prime=0.03$, where the Lyapunov exponent for the uncoupled system is zero.

\begin{figure}  
\centering
\scalebox{0.55}{\includegraphics{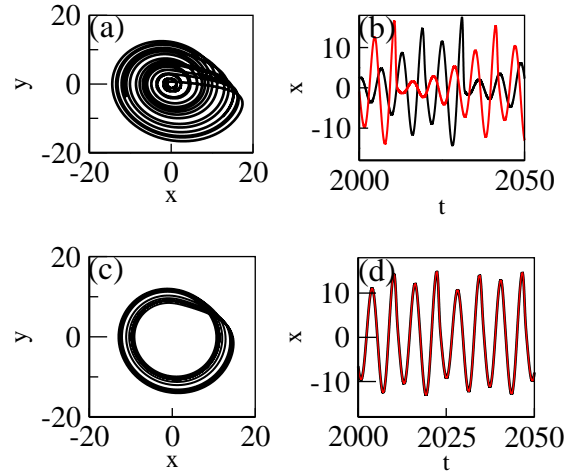}}
\caption{(a) The attractor of a typical oscillator from desynchronized group (b) time series of two desynchronized oscillators in the ensemble, (c) attractor of the oscillator from synchronized group and (d) time series of two synchronized oscillator.}
\label{fig3}
\end{figure}

 Thus, we see that the chimera state consists of distinct subpopulations corresponding to the different attractors. To describe the attractors in various groups, it is convenient to examine the projection of the attractor on the $x-y$ plane. As shown in Fig.~\ref{fig3}(a), the motion of the oscillators of desynchronized group is on a chaotic attractor which can be seen in the time evolution of the $x$ variable of two oscillators shown in Fig.~\ref{fig3}(b). We have also observed that some oscillators go to periodic  attractor but this is observed only when we explore their basin of attraction in Sec.(\ref{basin1}). Figure~\ref{fig3}(c) shows the projection of the dynamics in the $x-y$ plane for synchronized oscillators which can be inferred by the time series of the $x$ variable of the two oscillators as shown in Fig.~\ref{fig3}(d). \\

We also make use of qualitative measures based on the standard deviation of the nearby variables described in \cite{gopal,ghosh}. The state variables given in Eq.~\ref{eq1}, can be transformed as ${\bf u}_i={\bf x}_{i+1}-{\bf x}_i$. The total number of oscillators can be divided into $M$ (even) bins of equal length $n=N/M$ and the local standard deviation of the transformed state is given by
\begin{equation}
 \sigma_l(m) =  \left \langle \sqrt{ \frac{1}{n}\sum_{i=n(m-1)+1}^{mn}  \left[u_{l,i} - \langle u_l \rangle \right]^2} \right \rangle_{t} 
\end{equation}
 The quantity, $\sigma_l(m)$ is calculated for every successive $n$ oscillators. Thus, a measure Strength of incoherence (SI) is given by
 \begin{equation}
 SI=1-\frac{\sum_{i=1}^{M} s_m	}{M},~~~ s_m=\Theta(\delta-\sigma_l(m))
 \end{equation}
 where $\Theta$ is the Heaviside step function, and $\delta$ is a small predefined threshold. One observes that $SI=1$ for incoherent state, $SI=0$ for coherent state and $0<SI<1$ for the chimera state.

 This measure can be extended further to explore about the nature of the chimera state by introducing
 \begin{equation}
 \eta=\frac{\sum_{i=1}^M |s_i-s_{i+1}|}{2},~~~~(s_{M+1}=s_1)
 \end{equation}
 where, $\eta$ is $1$ for chimera states and for multichimera state $\eta$ is a positive integer value greater than $1$. For coherent states and incoherent states $\eta$ is zero.

\begin{figure}[htp]  
\centering
\scalebox{0.5}{\includegraphics{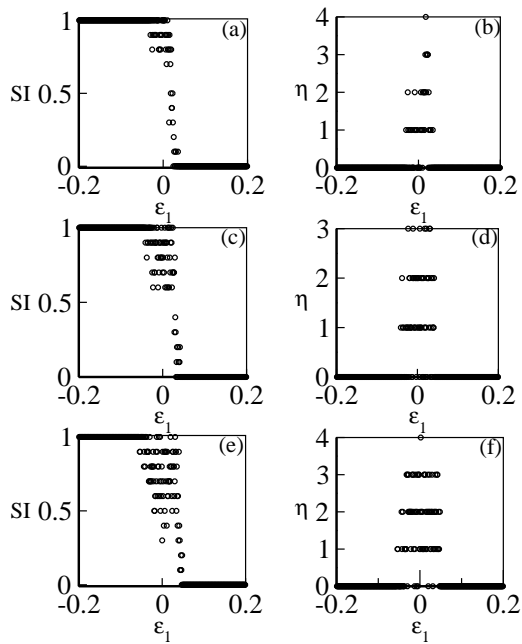}}
\caption{Strength of incoherence (SI) and the discontinuity measure $\eta$ are plotted  with respect to the coupling strength $\ep_1$ keeping the other coupling constant fixed at $\ep_2=0$ as shown in (a) and (b) respectively. (c) and (d) respectively shows the variation of $SI$ and $\eta$ for $\ep_2=0.1$ while in (e) and (f), SI and $\eta$ are respectively plotted for the $\ep_2=0.3$.}
\label{si}
\end{figure}
 
In Fig.~\ref{si}, we show the variation of strength of incoherence (SI) and discontinuity measure $\eta$ with $\ep_1$ for different values of $\ep_2$. It can be seen that by tuning $\ep_2$, one can effectively control the values of $\ep_1$ for which the occurrence of chimera state takes place in the system. As shown in Fig.~\ref{si} (a) ($\ep_2=0$), chimera states are observed at smaller values of $\ep_1$, as compared to Fig.~\ref{si}(c) where $SI$ is plotted for $\ep_2=0.1$. This observation is found to be consistent  for  $\ep_2=0.3$ case also that is described in Fig.~\ref{si}(e). These results can be justified by looking at the variation of the discontinuity measure $\eta$ for $\ep_2=0,0.1,0.3$ shown in Figures \ref{si}(b,d,f) respectively. Thus, one can infer that on increasing the value of the coupling parameter $\ep_2$ , one requires larger values of $\ep_1$ to observe chimera states.

\begin{figure}[htp]  
\centering
\scalebox{0.35}{\includegraphics[angle = 270]{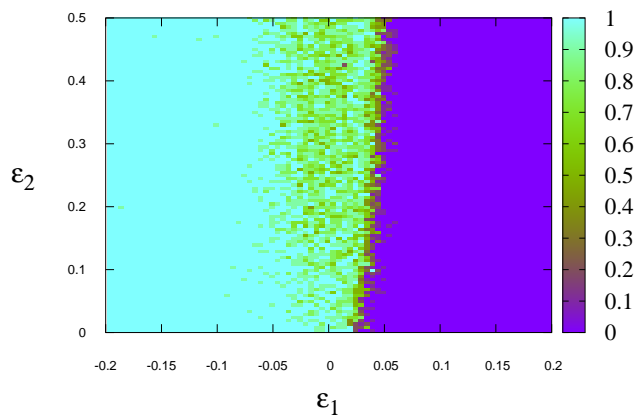}}
\caption{Strength of incoherence in the $\ep_1-\ep_2$ parameter space at $a=b=0.1$ and $c=9.0$ for an ensemble of globally coupled oscillators.}
\label{fig5}
\end{figure}

 One can also explore the entire parameter space $\ep_1-\ep_2$ to unravel the regions where one can find chimera states. We have plotted the Strength of Incoherence (SI) by simultaneously varying the coupling parameters as shown in Fig.~(\ref{fig5}). One can see that this phenomenon is generic in the sense that there exists a large region in the $\ep_1-\ep_2$ space where $0<SI<1$. As discussed earlier, $SI=1$ gives completely desynchronized state whereas $SI=0$ for synchronized state. Any intermediate value of $SI$ indicates a chimera state. Starting from zero along the positive $x$ axis, one reaches a region where all the oscillators are synchronized. This is the region where all the oscillators of the ensemble will be synchronized and hence chimera states can be avoided for all the values of $\ep_2$. Similarly, if one moves along the negative $x$ axis, a region is obtained where all the oscillators in the ensemble are desynchronized. Thus, our very choice of coupling in two variables is justified in the sense that we can now tune the coupling parameters ($\ep_1$ and $\ep_2$) in such a way that the coupled population may be restricted to a particular state of our choice. Thus,  all oscillators will be  synchronized in the region towards right, if one decreases $\ep_1$ below zero we enter a domain where all the oscillators are desynchronized. 

 Though the entire study so far was made only along the line $\ep_2=0.1$, but Fig.~\ref{fig5} strengthens our claim that the chimera states obtained using this technique are very general and can be obtained over a large area in the parameter space. It is also worth emphasizing that as we increase the value of parameter $\ep_2$, the range of $\ep_1$ where chimera states are observed can be controlled. Thus, one can suitably fix the value of coupling parameter $\ep_2$ such that the chimera state is obtained for selected values of $\ep_1$.

Our interest in the present work focuses on exploring the parameter space where chimera states can either  be created or destroyed for the correct choice of the coupling parameters. We  show that  chimera
states  generically  emerge in a rather simple network of globally coupled oscillators. So far our results suggest that if we couple two systems then we can selectively target bifurcation parameters such that the effective bifurcation parameter of the coupled system $a^\prime$ and $c^\prime$ takes a new value. At these shifted values, the dynamics, in case of autonomous system, may be periodic or chaotic depending on the value of the LE leading to synchrony or desynchrony in the coupled case.
 Remarkably, in our model the emergence of a chimera state depends mainly on tuning the parameter $\ep_1$ at fixed $\ep_2$, and the two parameters of the  system can be well adjusted  in  real  experimental  setup. Moreover,  there  is  a sufficiently  broad  parameter  range  where  the  chimera states exist. 
Though one can see the coexistence of synchronized and desynchronized states, but formation of chimeras can be explained only by looking at the basins of attraction. 

\begin{figure}[htp]  
\centering
\scalebox{0.5}{\includegraphics{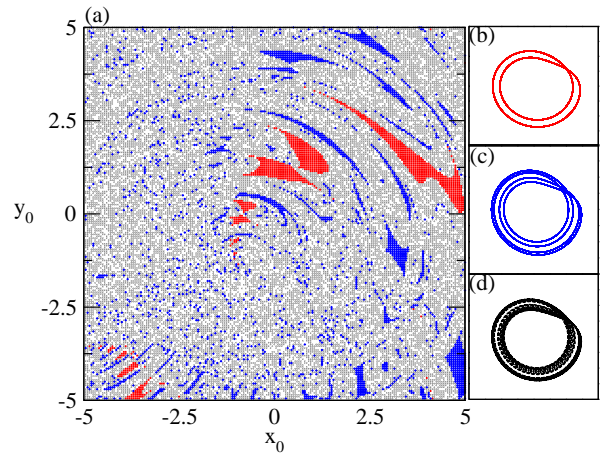}}
\caption{ Basin of attraction for two coupled oscillators for $a=b=0.1$, $c=9$, $\ep_1=0.023$ and $\ep_2=0.1$, where red (light grey) region corresponds to the period-2 attractors, initial conditions for period-4 are marked in blue color (dark grey). The black region in the basin of attraction are the initial conditions corresponding to the oscillators in synchronized group while the white region signifies the initial conditions for the desynchronized oscillators. (b) Typical period-2 attractor, (c) period-4 attractor and (d) typical attractor from synchronized group.}
\label{fig6}
\end{figure}

\section{Basin of Attraction and Lyapunov exponent}
\label{basin1}

 By appropriately tuning parameters, the system may be transformed from completely desynchronized state  to the mixed state which is very likely to be chimera to completely synchronized state. To verify whether a given state is chimeric or not we explore the changes in the basin of attraction of two mutually coupled R\"ossler oscillators with changing parameter values.  As shown in Fig.~\ref{fig6}(a), the basin of attraction for two coupled oscillators is riddled.  At $\ep_1=0.023$ the basin is completely interwoven in a complex manner and is completely intertwined for large volumes and it is very likely that the basin will be even more complicated for larger $N$. In general, the basin structure of different attractors in coupled systems is complex \cite{camargo, ujjwal}. Thus there is finite probability that two randomly selected nearby initial conditions will asymptote to different regimes that may be synchronized or desynchronized. Many studies have been devoted to describe the role played by the initial conditions in the creation of chimera states. It has been argued that chimera states may occur for random \cite{maistrenko, ujjwal} or  quasirandom \cite{showalter} initial conditions . 
 
 Another important region that is shown in Fig.~\ref{fig6}(a)  are the initial conditions that are period two (red or light grey) or period four (blue or dark grey) for which the typical attractors are shown in Fig.~\ref{fig6} (b) and (c) respectively. If two oscillators go to any of these groups then we observe that the two oscillators are not completely synchronized and hence they form part of the desynchronized group. However, it is worth mentioning that they are in phase synchrony with each other and hence we can characterize the chimera here as the one that exhibits groups where oscillators are in complete synchrony shown in black color in Fig.~\ref{fig6}(a), complete desynchrony described by white region in Fig.~\ref{fig6}(a) and a group that shows phase synchronization.

 \section{Stability of the chimera states: The Master Stability function}
 \label{msf} To determine the stability of the synchronous states, we apply the formalism of the Master Stability Function (MSF) that can be calculated based only on the knowledge about the dynamics of individual oscillators and the coupling function \cite{pecora-msf,huang}. A typical network of $N$ coupled oscillators  can be written as  $\frac{d{\bf x}_i}{dt}={\bf F(\bf x_i)}-\ep \sum_{j=1}^NG_{ij}{\bf H}({\bf x}_j)$, where ${\bf H(x)}$ is a coupling function, $\ep$ is a global coupling parameter, and ${\bf G}$ is a coupling matrix determined by the connection topology. The variational equations governing the time evolution of the set of infinitesimal vectors about the synchronous solution $\frac{d\delta{\bf x}_i}{dt}={\bf DF(s)}\cdot\delta {\bf x}_i-\ep \sum_{j=1}^{N}G_{ij}{\bf DH(s)}\cdot\delta {\bf x}_j$ that leads to the generic form of all decoupled blocks given by 
 \beq
 \frac{d\delta{\bf y}}{dt}=\left[{\bf DF(s)}-K{\bf DH(s)}\right]\cdot\delta {\bf y}.
 \label{eqmsf}
 \eqn
  \begin{figure}[htp]  
\centering
\scalebox{0.6}{\includegraphics{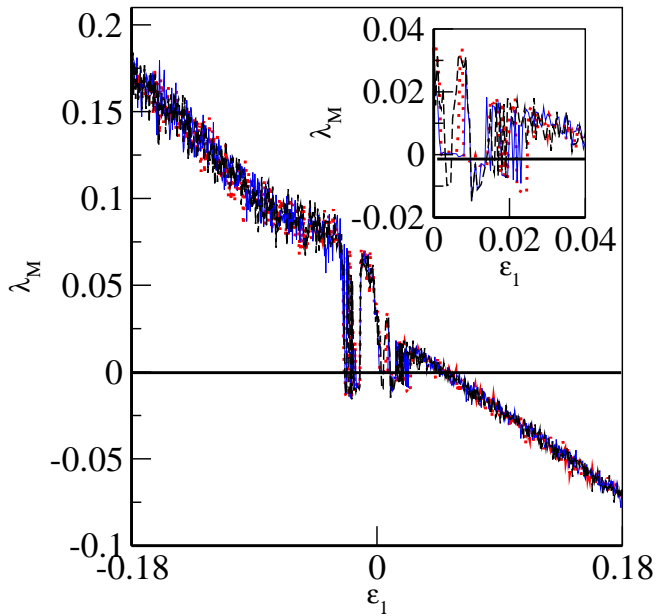}}
\caption{Variation of the Master Stability Function (MSF) corresponding to Eq.~\ref{eq1} with $\ep_1$ at $a=b=0.1$, $c=9$ and $\ep_2=0.1$ for different attractors described in Fig.~\ref{fig6}. The inset shows the region where we find different values for the MSF. }
\label{fig7}
\end{figure}

  The largest Lyapunov exponent for this equation $\lambda_M(K)$ gives the MSF. While the MSF describes the linear stability of the synchronous motion for a given attractor dynamics, in the presence of multistability it is important that one should rather look at the MSF of individual attractors. MSF corresponding to different attractors has been studied for R\"ossler like coupled oscillators and it was observed that the basin of attraction is complex leading to diverse dynamics. \cite{bocaletti}.
 For coupled R\"ossler flows given by Eq.~\ref{eq1}, we fix $\ep_2=0.1$ and calculate $\lambda_M(\ep_1)$. It can be seen that the MSF becomes negative as the value of $\ep_1$ is increased implying stable synchronized dynamics as shown in Fig.~\ref{fig7}. However, very close to the zero crossing, is the region of chimera states \ie mixed region having both coherent and incoherent dynamics. In Fig.~\ref{fig7}, we have plotted $3$ curves corresponding to $3$ different attractors that have already been described in Sec.~\ref{basin1} (Fig.~\ref{fig6}). As shown, for small value of $\ep_1$, MSF is positive implying that the population of the ensemble will be desynchronized. As one increases $\ep_1$, $\lambda_M$ becomes negative indicating that the oscillators are synchronized. Interestingly, it can be seen that $\lambda_M$  has different behaviour for different attractors, this implies that for a given value of $\ep_1$, the dynamics over a given attractor may be synchronized, while for other attractors it may not be synchronized. This results in a state where we have a coexisting coherent and incoherent regions. In Fig.~\ref{fig7}, the inset clearly indicates that at around $\ep_1=0.02$, MSF corresponding to one of the attractor is negative (red dotted line) while the MSF corresponding to the second attractor is complete line (blue curve), while  the completely positive (dashed black line) corresponds to the third attractor. Hence, the MSF being positive and negative simultaneously for different attractors indicates a state where one can have coexisting coherent and incoherent regime.

\section{summary} 
\label{summary}
In this work, we have proposed a new scheme that will lead to the emergence of dynamical chimeras in the ensembles of coupled chaotic oscillators. This is possible in the absence of explicit nonuniformity in the coupling, time delay and multistability unlike the previous studies where chimera states were observed in the presence of any of these. We considered an ensemble of globally coupled R\"ossler oscillators and by tuning the coupling constants one can create chimeras with desired features. By further tuning these parameters, one may create a completely synchronized or desynchronized state indicating  that the chimera states can be created for a given set of parameter values or can be avoided  for other values. It has also been observed that the basin of different attractors are intertwined in a complex manner indicating that it is impossible to avoid the chimera states. It is very likely that the two nearby initial conditions will go to different attractors. The transition from completely synchronized state to an incoherent state via a mixed state has been verified by employing quantitative measures namely the strength of incoherence and discontinuity measure. 

Robustness and stability of these dynamical states can be verified with the help of Master Stability Function which is negative when the entire population of the ensemble is synchronized and positive if it is desynchronized. However, very close to the transition point where we have observed chimera states, it is clear that the MSF behaves differently for different attractors. Negativity of the MSF indicates that the synchronized dynamics is stable and otherwise it is positive. Coexistence of negative and positive values of the MSF is a clear indication that for these coupling parameters, coherent and incoherent states can coexist. Thus, MSF for multistable systems provide information about the synchronizability of the network that helps us understand how the modification of coupling parameters leads to the synchrony or desynchrony of the oscillators. It also provides a generic tool for exploring the complete dynamics in the ensemble.

Our results show that by fixing one of the two coupling parameters say, $\ep_2$,   it is possible  to shift the population of oscillators to a state that is apriori required. In a power grid, the generators are synchronized. Under the influence of perturbations the synchrony may be fully or completely destroyed \cite{moter,dorfler}. Thus, it is important to identify tunable parameters in the system that ensures synchrony of generators. In brain, extended periods of synchronization are pathological \ie it is a symptom of seizure \cite{arthuis}. In many such systems it is not possible to change system parameters, therefore one can adopt the techniques outlined in \cite{pisarchik} to bring the effective parameter to the required state.
  
\section*{Acknowledgment:} AAK acknowledges UGC, India for the financial support. HHJ would like to thank the UGC, India for the award of grant no. F:30-90/2015 (BSR). We also thank Awadhesh Prasad for useful discussions.

\end{document}